\documentclass[12pt]{article}
\newif{\ifMarginalComments}
\MarginalCommentstrue
\usepackage[a4paper,margin=2.7cm]{geometry}
\usepackage{latexsym,pifont,color}
\usepackage{amsmath}
\usepackage{amssymb}
\usepackage{pslatex}

\newcounter{ncomm}

\begin{document}

\title{Modelling the Evolution of Programming Languages} 
\author{Silvia Crafa  \\ Universit\`a di Padova}
\date{}
\maketitle

\begin{abstract}
Programming languages are engineered languages that allow to
instruct a machine and share algorithmic information; they have a
great influence on the society since they underlie almost every
information technology artefact, and they are at the core of 
the current explosion of software technology. 
The history of programming languages
is marked by innovations, diversifications, lateral transfers and
social influences; moreover, it 
represents an intermediate case study
between the evolution of human languages and the evolution of
technology.

In this paper we study the application of the Darwinian explanation to
the programming languages evolution by discussing to what extent the
evolutionary mechanisms distinctive of biology can be applied to this
area. 
We show that a number of evolutionary
building blocks can be recognised in the realm of computer
languages, but we also identify critical issues.
Far from being crystal clear, this fine-grained study shows to be a
useful tool to assess recent results about programming languages
phylogenies. 
Finally, we show that rich
evolutionary patterns, such as co-evolution, macro-evolutionary
trends, niche construction and exaptation, can be effectively applied
to programming languages and provide for interesting explanatory
tools.   
 

\end{abstract}

%%%%%%%%%%%%%%%%%%%%%%%%%%%%%%%%%%%%%%%%%%%%%%%%%%%%%%%%%%%%%%%%%%%

\section{Introduction}

The Darwinian theory of evolution has been often applied to cultural
  systems, both to model the development of specific cultural traits
  and to provide a general explanatory framework. 
One main question raised by the literature is about how deep is the
analogy between biological and cultural evolution (e.g., 
\cite{ClaidiereEtAl,LaneEtAl}). Variation, selection and inheritance
often operate very differently in the biological and cultural cases,
and very different cultural traits, like for instance artefacts in the
material cultural world and the cultural propagation of behaviours,
arguably require different explanations.

In this paper we focus on the evolution of Programming
Languages (PLs), a specific aspect of software systems that represents
an interesting case study, lying in the intersection between
two notable streams of works in the realm of cultural evolution: the
evolution of human languages and the evolution of technology. 
Programming languages have a great influence on the society, since
they underlie almost every information technology artifact.
%; they even inform smart tags that are ubiquitous in real life. 
Moreover, the PL arena is very lively, crowded and dynamic: there
continuously appear new languages, mainstream languages strongly
compete, often pushed by companies that employ such languages in the
technology they sell, and even measuring the success of a PL is a
very difficult and debated task \cite{LangPop}.

Like human languages, PLs are a means of
communication: they allow a programmer to instruct a machine, they
allow two or more programs to interoperate, 
and they also allow programmers to share algorithmic,
i.e, formal and precise, information. 
The first programming languages date back to the 1950s; their history
might seem short compared to other cultural systems, but it is very
fast and rich, marked by innovations, diversifications,
lateral transfers and social influences.
Compared to biological systems, variations in both human and
programming languages are affected by intentional choices. 
Intentionality is particularly sharp in the case of PLs, 
which are carefully designed for a specific goal, and they are the
result of a planned combination of elements used in other
programming languages, according to the tinkering practice distinctive
of the technological evolution.
However, even if PLs locally evolve according to a planned design, 
%(not to say ``intelligent design''...), 
the macro-history of PLs has been clearly affected by a number of
mechanisms distinctive of sociocultural systems;
modelling their complex evolution is of critical importance in order
to analyse the current explosion of software technology.

We observe that programming languages provide for a well defined and
expressive realm, that represents a
interesting subject for studying and testing evolutionary patterns,
since it shares commonalities with other cultural and
technological systems, and at the same time it displays specific
features. 
We then study the application of the Darwinian explanation to the PLs
evolution, trying to unveil evolutionary patterns and driving forces
that guide, or unfold behind, the development of this rich
scientific area. 
However, we stress the fact that, rather than casting PLs evolution
into the Darwinian account, our main goal 
is to discuss to what extent the evolutionary mechanisms distinctive
of biology can be applied to this area, shedding some light on how much
the rich Evolutionary Theory's research program can provide for an
explanatory framework, still calling for a pluralism of
explanations.

\medskip
More precisely, we start in Section 2 by describing the aspects that
distinguish the evolution of technology and that of human languages
from the biological evolution. We then point out that in order to
specifically address the case of programming languages, it is
important to precisely understand to what extent the building blocks
of the evolutionary explanation can be rephrased in this context. 
Taking inspiration from Pagel's table detailing the parallels between
biological and linguistic evolution \cite{PagelNature}, we extend
in Section 3 such a parallel by discussing how far the
basic ingredients of the Darwinian evolution can be rephrased to deal
with PLs. Far from being crystal clear, the rough identification
of the building blocks of the PLs evolution shows to be a useful  
tool to assess the results obtained by R. Sol\'e and S. Valverde
\cite{SoleValverdePLEvo}, who applied systematic phylogenetic
methods to infer 
evolutionary trees of programming languages starting from a network
of influence. Finally, we show in Section 4 that richer evolutionary
patterns, such as co-evolution, macro-evolutionary trends, niche
construction and exaptation, can be effectively applied to 
programming languages, and provide for interesting and useful
explanatory tools.

\section{Programming languages: an intermediate case between the evolution of technology  and that of human languages}

\paragraph{The evolution of technology.}
The comparative study of technological development and biological
evolution is a very rich and lively topic: the
analogies and differences are so profound that delving into this
comparison is still a source of insightful thoughts. 
%In this paper we
%specifically address information technology and software systems 
Like biological evolution, the technological development displays 
a process of descent with variation and selection, which includes
convergence, contingency and extinction. Even an elaborate evolution
pattern like punctuated equilibria applies also to the technological
progress: see \cite{LochHuberman} for a detailed discussion about
the existence of punctuated equilibria in technology diffusion. 
 At the same time, differently
from biological systems, technological innovations are examples of
\emph{planned design}, ranging from short-term goals (e.g. a safer
car) up to 
long-term expectations (e.g. ubiquitous computing initially achieved
by means of laptops, then by means of smartphones, now by the Internet
of Things scenario). In particular, the leading role of planned design
is here especially magnified by the fact that technology often offers
a clear notion of 
\emph{measurable progress} to be reached, in terms of efficiency, 
correctness, safety, cost or speed.

It is also well known that \emph{tinkering}, that is the widespread
reuse and combination of available elements, is a typical feature of
the evolution process. New technologies often emerge as a
recombination of preexisting technologies, in a similar way as new
biological structures reuse available elements. 
However, as observed in \cite{EvoTech}, the impact
of introducing new simple technological elements can be very high,
and completely reset the path of future technologies, whereas in
biology established solutions are seldom replaced.
On the other hand, a crucial point is that
the study of technological innovations, and in particular
that of information technology, must significantly take into account
the \emph{interactions with social end economic factors}. Nowadays
information technology, economy and social systems are deeply
interconnected and interdependent, each one being able to transform
the development landscape of the others. We claim that it is not
just a matter of mutually affecting, co-evolving, domains, but 
these days information technology, economy and social systems can be
better interpreted as a proper \emph{ecosystem}.
For instance, in the case of information technology, issues of
retro-compatibility, but also market dominance or trends, often limit
or make impossible the spreading of better solutions, while the
dominant technology keeps stuck to suboptimal products. As an
example, Web-based solutions for software applications are sometimes 
dictated by a trend, whereas classical client-server architectures
avoiding browsers would have been best suited. Moreover, technological
innovations like the Internet, Cloud computing and Big Data, have
been so impactful on the society, that they are no more just
scientific words, but they are also economical and social keywords.
%Finally, the US programs aiming to boost the computer science
%education will inject new assets that will likely infulence the
%development of computer science and economy later.

\medskip
One of the pivotal components of information technology
artefacts is represented by software systems, which essentially aim at
controlling the behaviour of physical components. 
Interestingly, software provides a particularly well
preserved fossil record, hence it is a good candidate to 
re-apply
%test the applicability of 
quantitative analysis methods, such as phylogenetic
relationships reconstruction, which are well established in
evolutionary biology. 
However, the applicability of these methods for the 
%phylogenetic methods have not been substantially applied to 
reconstruction and the analysis of software evolution raises a number
of issues. First of all, as observed by R. Sol\'e et al. in
\cite{EvoTech}, 
software systems offer \emph{multiple levels of detail}:
the code written in a given programming language, the architecture of
interacting pieces of code and data, the social network of
engineers that designed the software system and those that maintain it
over its life-cycle. In this setting even appropriate definitions of
what would correspond to mappings between genotype and phenotype is
far from being trivial.
The second issue comes from the observation that \emph{the phylogeny
  of technology is not hierarchical}: the combinatorial effect of
tinkering, 
the rapid and large information exchange, the ageing process of
technologies, entail reticulate phylogenies similar to that of
bacteria. Then reticulate networks, instead of trees, appear to be
more appropriate when dealing with technological innovations.
The study of patent networks 
provides a clear example of this phenomenon: it shows instances of
different patterns like gradual evolution, stasis followed by
punctuation, extinction, selection and  even resurrection
(\cite{EvoTech,VSBP}). Moreover, the multi-parental genealogy of 
patented inventions calls for highly multi-parental, possibly
multilevel, lineages of evolution (\cite{Bedau}).

\paragraph{The evolution of human languages.} Human languages are 
well recognised culturally transmitted replicators. They put forward 
multiple ways in which individuals learn from one another, and
give evidence of how cultural traits get distributed through the
different channels of social transmission.
The study of their complex evolution provides insights about 
the interplay between biological and cultural evolution, and
highlights the role of different evolutionary mechanisms.
The recent debate
(e.g. \cite{ClaidiereEtAl,Suman,ScottPhill,AcerbiMesoudi,Kovaka}) 
emphasizes 
the disanalogies between the way variation, selection and inheritance
operate in the biological and cultural cases and 
calls for a generalisation of the classical Darwinian selectional and
replicative models. 
Indeed, cultural transmission displays both preservative
and constructive transmission aspects; for instance,
while propagating the word sound can only rely on copying, propagating
the word meaning triggers constructive, re-productive, processes
\cite{ClaidiereEtAl}. Moreover, phenotypic plasticity, developmental
constraints, niche construction and inclusive inheritance have been
pointed out as major factors that, together with selection, shaped
the complex traits of the human language \cite{Suman}.

In the next section we will argue that programming languages similarly
entail peculiar mechanisms of transmission, and they also possibly
require a generalisation of the Darwinian replicator model.
Moreover, in this context the variation and selection processes are
not causally independent; it is often the case that the very same
people have a role both in the design of new languages (i.e., in
variation) and in determining a language success or failure (i.e., in
selection). This is just an example of the sociocultural effects on
the PLs development, and it is reminiscent of 
Lane et al.'s reciprocality principle: ``the generation of new
artefact types is mediated by the transformation of relationships
among agents/organisations; and new artefact types mediate the
transformation of relationships among agents/organisations''
\cite{LaneEtAl}.

\medskip
An interesting setting for comparing the evolutions of human languages
and programming languages is the study of their phylogenetic
history. By systematically applying well developed techniques of
phylogeny reconstruction, and cautiously assessing their results, rich
and nontrivial dynamics may be identified. 
However, this approach is full of methodological and epistemological
issues.
When comparing human languages and PLs, a first important remark is
about the topology of language phylogenies. 
In \cite{MHPhyloCulture} the authors observe that
the phylogenetic analyses of human languages show that trees are 
well suited models for describing language histories, even if they
evolve both vertically and horizontally. This can be explained by
observing that human language phylogenies have been constructed on top
of a fundamental vocabulary, that is a data set made of 200 words
(Swadesh list) corresponding to conservative, cross-culturally
universal meanings, such as 'mother' or 'sun'. These words tend to
evolve slowly and resist to lateral influences, they can be seen
as adaptive cultural traits,%\SC{si puo' dire?}, 
that hence more
likely have a phylogenetic (vertical) signature. Moreover, families of
human languages, such as Indo-European and Bantu languages do not
hybridise so much, which also explains why human language phylogeny is
shaped like a tree with monophyletic groups.

The case of programming languages is different and particularly
subtle. 
First of all, the methods applied to the study of human languages
cannot be directly applied to PLs: defining a sort of
Swadesh list for PLs would be very controversial, if not meaningless. 
In \cite{SoleValverdePLEvo} systematic phylogenetic
methods have been applied to infer evolutionary trees of PLs starting
from a network of influences. However, the nature of
technological innovations described above suggests that the results in
\cite{SoleValverdePLEvo} might underestimate the reticulated nature
of PLs phylogeny. We will more precisely discuss the approach of
\cite{SoleValverdePLEvo} in the next section, but we point up %here
that  prior to phylogeny reconstruction, lay unsolved questions like
what should a PL phylogeny account for?
and what is the ``adaptive core''%\SC{si pu\'o dire?} 
of a programming language?
We think that the first step to answer these questions is to precisely
study the basic characteristics of programming languages so to
understand to what extent they can be effectively cast into the
evolutionary framework. 
%the building blocks of their evolutionary explanation.
%, and possibly isolate,

%{\bf DISREI DI METTERE QUI LA PARTE SULLA RICERCA COME CONTROLLO DI MUTAZIONI.....}

\section{The basic ingredients of PLs evolution}
%What is a PL? In what sense does it evolve?
%\begin{itemize}
%\item 
%Discuss the syntax vs the semantics: {\bf what is a species?}
%\item 
%Highlight the case of programming paradigms as {\bf group/class}
%\end{itemize}

%\medskip
In \cite{PagelNature} Mark Pagel gathers into a table a detailed 
parallel between biological and linguistic evolution. The same table 
is adapted and extended in \cite{MHPhyloCulture} to deal with
cultural systems. In this section we virtually add a column to such a
table, extending the parallel to encompass the case of programming
languages.

As observed by Pagel, the key among these parallels is that both
biology and human languages are digital systems of replicators, that
is they both include discrete heritable units, respectively the
genes and the words or phonemes, that assure transmission fidelity. 
In the case of PLs, discrete units correspond to language
primitives, such as loops, conditional commands, functions, but also
objects or threads in richer languages.
Another crucial aspect is the identification of
the replication mechanisms: in biology inheritance comes essentially
with parent-offspring, while human language is replicated by means of
teaching, learning and imitation. 
%As we reported in the previous section, a stream of
%literature argues that in the case of cultural systems relying on
%replicators and preservative mechanisms as the only source of
%powerful 
%inheritance is not satisfactory, and puts forward a number of
%additional transformative and constructive mechanisms. 
For programming languages there is
no filiation mechanism, programmers learn a coding language and assess
their knowledge by checking the execution of the programs they write.
It is important to observe that in the learning phase the language
cannot be modified, in particular 
there is no pragmatic semantics for PLs: a program
will not be executed unless it fully respects the required syntax.
For instance, even missing a simple symbol like a  curly bracket can
cause a C++ program to be rejected by the machine.
%Imitations can be recognized between different languages, but not in
%the learning process. 
Replication and inheritance processes can also characterise the
life-cycle of a given piece of software, but this aspect is rather
related to the software engineering practice, that corresponds to a
different observation level, that is orthogonal to our focus on the
coding language. 
% of software systems that is different from the one
%we are interested in here.

%
In order to give a precise account of inheritance and a 
number of related concepts, we cannot avoid trying to clearly
delineate our 
observation level by addressing the pivotal issue of defining what is
a programming language species.%\SC{Per HL c'e' definizione di specie??} 

\paragraph{What is a PL species?}
In analogy with human languages, it is rather natural to say that a
species corresponds to a distinct programming language, such as
Fortran, Java or C++. The analog of people talking the same language
is instead less clear: individuals belonging to the same species, say,
``the Java species'', might correspond to the programs written in
Java or to the Java programmers. We claim that ``populating'' a
programming language with people that code in that language shades
the distinctive features of computer languages. For instance,
programmers are usually polyglot, but they do not mix programming
languages: multi-language software systems are precisely defined in
terms of separate mono-language modules that interoperate according to
a precise algorithm, that is in turn implemented in another
formal language.
On the other hand, letting the individuals correspond to the programs
brings in a number of deep consequences. 

First of all, we remark that each programming language comes 
with a \emph{language specification} that precisely defines which
program phrases belong to that language: different
Java programs may behave differently or use different subsets of
Java primitives, but it is impossible for a program to use an
altered version of a Java construct or a new piece of syntax, since
the machine would recognise it as an error, that is an ill-defined,
non-executable program. As a consequence, a language specification
provides for a clear definition of the species boundaries, but at the
same time in this view 
we have that PL species provide no individual variability, therefore
the very 'population thinking'\footnote{
``Population thinking involves looking at a system
as a population of relatively autonomous items of different types with
the frequency of types changing over time'' (\cite{ClaidiereEtAl}).}
, which is the essence of the
evolutionary framework, seems to fade. 

To be precise, the language specification is not always so sharp and
accurate: 
for instance, the syntax of the C language has a standard definition
in terms of a formal grammar \cite{ANSIC}, 
while the Java specification language is given as
a written English document \cite{JavaSpec}. 
In some case it is difficult to trace the
boundaries between a language and its libraries (e.g. Python) or there
might be just a reference implementation with extensive test
suites instead of a language specification, as for Perl. 
Nevertheless, for any programming language, in order to execute a
program written in that language, the computer has first to translate
that program into machine instructions. Therefore each PL
requires an automatic translator (a compiler or an
interpreter) that must be designed according to some parsing rules
that recognise syntactically well-defined programs. Hence the
specific syntax definition instructing the language parser happens to
determine the species boundaries.
In particular, programming languages \emph{do not hybridise}: a
program mixing Java commands and C++ commands is rejected by the
compiler. Hybrid languages might still be defined, they are
often called dialects; however their definition must encompass the
dialect's parser with a precise specification of which program
is recognised to belong to the dialect and which is not. Henceforth
the dialect is not an hybrid but it is another species itself.
%\footnote{An alternative could be let an individual correspond to
% the set of instructions obtained as the result of 
% the translation of the program into machine commands. Such a
% low-level definition of individual entails the individual
% variability, since the translation is hardware-dependent; 
% however it seems to be inconsistent with the identification of 
% syntactic primitives as the discrete heritable units. {\bf questo lo cancellerei}}.

Therefore we have that, %, interstingly, 
on the one hand, even if the
language specification may be blurred in some case, 
%especially for new languages that have not been standardized yet,
%the important point here is to highlight that the 
the definition of PL species seems to be sharper than that, still very
controversial, of human languages species and biological species. 
On the other hand, by losing individual variability the populational
explanatory framework can hardly be applied. 
Nevertheless, despite this crucial difficulty, which we highlight as
an open problem, we think that a number of evolutionary building
blocks can still be recognised in the realm of computer languages.

\begin{table}[h]
{\small
\begin{center}
\begin{tabular}{|c|c|c|}
\hline
{\bf Biological Evolution} & {\bf Human Languages Evolution} & {\bf PLs Evolution}
\\[5mm]
{\it Discrete heritable units} \hfill \mbox{ } & & \\
\hline
nucleotides, aminoacids, genes & words, phonemes, syntax &
                         primitives, syntax
\\[5mm]
{\it Mechanisms of inheritance} \hfill \mbox{ } & & \\ 
\hline
reproduction, occasionally clone & teaching, learning, imitation &
                         fixed specification
\\[5mm]
{\it Hybridisation} \hfill \mbox{ } & & \\ 
\hline
species mixes & language Creoles & no: hybrid code does not execute
\\[5mm]
{\it Mutation} \hfill \mbox{ } & & \\ 
\hline
genetic alteration & new words, mistakes, sound changes & specification/version update
\\[5mm]
{\it Speciation} \hfill \mbox{ } & & \\ 
\hline
allopatric, sympatric, ... & lineage splits (e.g. geographic, & 
                                           major language update  \\
           & social, ethnic -separation) & Domain Specific Languages\\           &                             & e.g., Csound
\\%[2mm]
{\it Anagenesis} \hfill \mbox{ } & & \\ 
\hline
evolution without splitting & linguistic change without split & 
                                      just minor updates\\
                            & & e.g., PostScript
\\[2mm]
{\it Horizontal transfer} \hfill \mbox{ } & & \\ 
\hline
horizontal gene transfer & borrowing & lateral influence but no hybrids
\\[5mm]
{\it Clines} \hfill \mbox{ } & & \\ 
\hline
geographic clines & dialects and dialect chains & dialects are
different languages
\\[5mm]
{\it Drift} \hfill \mbox{ } & & \\ 
\hline
genetic drift & language drift & no random sampling effect
\\
\hline
\end{tabular}
\end{center}
\caption{Some analogies between biological, human languages and
  programming languages evolution.}
\label{tab:PagelTab}
}
\end{table}

\paragraph{Diversification processes.}
%Given the sharp notion of species above, Table~\ref{tab:PagelTab}
%states that the PLs replication mechanism is cloning. 
%On the other hand, a number of evolutionary concepts related to the
%species modifications can be observed also in the case of programming
%languages. 
Even if the sharp notion of species given above entails no individual
variability, we can recognise 
a language mutation when an updated language specification
is released. For instance, the Python 3.4.0 language would hardly be
considered a different language from Python 3.3.3. Moreover, issues
of backward-compatibility impose that the programs written according
to the old specification must be correctly interpreted also by the new
parser. In addition, whenever a program construct is abandoned in the
new release, 
the new language specification labels such a construct as a
``deprecated feature'', that is a primitive that can still be
correctly parsed (for backward-compatibility) but that must not be
used anymore. Interestingly, deprecated features could be assimilated
to \emph{vestigial traits} in biology.

Whenever the language specification undergoes a major update, a
speciation event occurs. For instance Java 8 is in many sense a
different language from Java 7 because it provides new language
constructs that deeply modify the programming
style\footnote{significant efforts have been made to guarantee
  backward-compatibility.}. Moreover, a parallel can be established by
biological allopatric speciation, that is,
speciation by means of geographic separation and diversification, 
and domain specific languages (DSLs), that is, PLs
specialised to particular application domains. For instance, Csound
is an audio programming language used by composers and musicians
derived from the C language. On the other hand, biological anagenesis,
that is, species evolution without splitting events, can be recognised
for PostScript, the language used by laser
printers, whose specification underwent only minor updates.

The final three rows in Table~\ref{tab:PagelTab} are direct
consequences of the sharp definition of PLs species and the fact that
PLs are explicitly/intentionally designed.
Lateral gene transfer and new words acquisition by borrowing have no
parallel in programming languages. There are certainly lateral
influences in the design (and update) of a language specification,
i.e., at speciation events, but not between individuals. Similarly,
geographic clines and dialect chains can be only superficially
compared with PL dialects: for instance a program written in a
dialect of Java is in general not recognised by the Java compiler,
thus it can directly interoperate with standard Java 
software only if it just relies on the common language core.
%\footnote{To be precise, a dialect language in general just
%  extends the core of the original language with new
%  constructs. Hence 
%  interoperability is generally guaranteed as long as it only 
%  relies on the common language core.}.
Finally, genetic and language drift have no parallel in the context of
programming languages, because there is no random sampling effect in
the replication mechanism.

\paragraph{External processes.} We complete our analogies in
Table~\ref{tab:PagelTab-2}, starting with the description of the
selection processes. In the case of programming languages, as well as
in general for technology, the natural selection distinctive of
biology can be compared to the inherent concept of scientific
progress. A PL 
 that allows a more efficient implementation, or that is safer of less
 error-prone, or that is targeted to a more advanced hardware, will 
survive other languages. On the other hand, social selection and
trends operate on PLs similarly to other cultural systems. 
As discussed in the previous section, strong selection pressures also
come from economic factors, that might overcome ``natural'' selection
and determine the dominance of sub-optimal solutions.

Language extinction happens also in computer science, for instance
in the case of low-level languages explicitly targeted to obsolete
hardware. 
The case of the Cobol language is curious: it is definitely an
obsolete language but most of the financial software systems consist
of stable Cobol programs and porting such systems to a different
language opens the way to the introduction of errors, maintenance and
compatibility issues that might be not sustainable. The adopted
solution is to keep the core software written in Cobol and wrap it
with a front-end written in a modern language. This solution reminds
of the \emph{canalisation} processes found in biology.%\SC{controlla se  giusto...}. 
Finally, as an example of announced extinction, we mention the
Objective-C language. It is variant of the C language that has been
designed in the 1980s but that became mainstream only in 2000 when
Apple imposed its usage to develop applications for its mobile
devices. In 2014 Apple released Swift, a modern programming language
that makes programming easier, safer and faster, thus actually
condemning Objective-C to extinction. Clearly, Objective-C will still
be necessary to maintain the applications written in Objective-C, but
mobile applications become obsolete very quickly, thus the language
will probably fade away.

\begin{table}[h]
{\small
\begin{center}
\begin{tabular}{|c|c|c|}
\hline
{\bf Biological Evolution} & {\bf Human Languages Evolution} & {\bf
  PLs Evolution} 
\\[5mm]
{\it Selection} \hfill \mbox{ } & & \\
\hline
natural selection & social selection and trends &
                         progress, social selection, trends
\\[5mm]
{\it Extinction} \hfill \mbox{ } & & \\
\hline
species (mass) extinction & language death & language death
\\[5mm]                                                         
{\it Fossils} \hfill \mbox{ } & & \\
\hline
fragmented fossil records & ancient texts & dead languages, deprecated
features
\\[5mm]
{\it Evolution rate and awareness} \hfill \mbox{ } & & \\
\hline
slow, not planned & fast or slow, partially intentional &
                         fast, fully designed
\\
\hline
\end{tabular}
\end{center}
\caption{Further analogies between biological, human languages and
  programming languages evolution.}
\label{tab:PagelTab-2}
}
\end{table}

The final row in Table~\ref{tab:PagelTab-2} illustrates the difference
in the evolution rates: biological species evolve very slowly, while
PLs evolution is extremely fast. Moreover, there is no intelligent
design guiding biological evolution, whereas every choice in the
specification of a programming language is intentional. It is
important to observe that even if local choices are carefully
designed, the global evolutionary process for PLs is not completely
planned: long-terms effects might have not been planned, and reactions
to social, economical or contingent factors can hardly be anticipated.
Some of these non completely intentional effects will be discussed in
the next section.

\subsection{The programming paradigms as a multilevel issue}
%{A critical review of Valverde-Sol\'e's paper: the case of
%  programming paradigms}

Recently S. Valverde and R. Sol\'e \cite{SoleValverdePLEvo} put
forward a quantitative study of the evolutionary dynamics displayed by
programming languages. They consider a dataset of 347 PLs, appeared
between 1952 and 2010; an influence graph is then reconstructed by
extracting from Wikipedia the list of PLs that influenced the design
of each PL. From such a graph, which is a quite tangled and complex
network, the authors extract a phylogenetic tree following the
approach 
of \cite{GualdiYeungZang} developed in the context of networks of
citation 
in scientific publications. More precisely, the method generates a
backbone based on identifying the most influential parent for each
language, and an additional graph that keeps all the horizontal
exchanges among languages.

The proposed method is ingenious since other quantitative approaches
rely on syntactic similarity measures that can hardly be defined for
PLs. The method finds two, disjoint and highly imbalanced, major
clades corresponding to the imperative and the functional programming
families, together with several smaller classes, thus accurately
mapping the known historic development of PLs. 
Moreover, the bundle of links observed in the horizontal transfer
graphs supports the combinatorial rule of technological evolution
mentioned above. On the other hand, we think that one %a couple of 
major authors' remark about the obtained results deserves a deeper
examination. 

%\paragraph{Punctuated pattern.}
%The authors observe that the evolutionary dynamics displayed by PLs,
%in particular the high imbalance of trees and the bundled horizontal
%interactions, 
%reveal the presence of non-uniform rates of change, with bursts of
%diversification events tied to innovations, thus supporting a
%punctuated pattern of evolution.\SC{Questo paragrafo \`e da scrivere} 
%{\bf Ricordo che Telmo era perplesso su questo: la presenza di grande
%  quantit\`a di ricombinazione/tinkering \`e in contrasto con il
%  pattern punteggiato....corretto?}
%\emph{``Using both data analysis and network modelling, it is shown that
%programming language evolution is highly uneven, marked by innovation
%events where new languages are created out of improved combinations of
%different structural components belonging to previous languages. These
%radiation events occur in a bursty pattern and are tied to novel
%technological and social niches. Our method can be extrapolated to
%other systems and consistently captures the major classes of languages
%and the widespread horizontal design exchanges, revealing a punctuated
%evolutionary path.''} {\bf Io non vedo nei loro grafici per forza un
%  pattern punteggiato, ma solo ogni tanto qualche linguaggio piu' di
%  successo/influente. Non vedo quali sono i ``radiation events'' ne'
%  si capisce davvero che sono in un ``bursty pattern''. Mi sembra
%  ragionevole che novel technologies and social niches producano
%  radiation events, ma non vedo come questo emerge dai loro grafici! }

%\paragraph{Object-Oriented Programming as a multilevel issue.}
The authors observe that in many of the lineages obtained by their
method there are examples of languages displaying object-oriented (OO)
traits. They claim that the historical separation of imperative and
functional programming transitions to a \emph{convergent evolution}
towards object orientation. Indeed, 
%OO traits are shared by many (lineages of) PLs essentially because 
OO-programming's strengths in abstraction,
modularity, dependency management and code reuse made this paradigm a
de facto standard for the development of reliable large-scale
software.   
However, we think that explaining the emergence of OO-programming
in terms of convergence is not fully satisfactory and biased, 
essentially coming as a consequence of choosing trees, rather than
networks, as a working model. As we discussed in the previous section,
trees might underestimate the reticulated nature of technological
phylogenies. 
We then propose a different explanation, based on a precise account
of the notion of programming \emph{paradigms}. 

Given a PL, a program
must be written according to the precise language syntax, however
there is some freedom about the adopted style, which marks 
the program's paradigm. A programming paradigm
(e.g., imperative, functional, object-oriented, declarative, logic)
is a set of programming patterns that characterise the
structure of programs and entail a fundamental style of computer
programming. In computer science paradigms are also used to classify
PLs 
into taxonomies. Using the biological talk, we can say that if a PL
is a species, a paradigm is a family or a class. However, differently
from biology, there is no unique, generally accepted, classification
(see 
e.g. \cite{WikiTaxo}) because paradigms cannot be formally defined as
the language syntax, and more importantly because there are aspects of
languages that do not neatly divide up into paradigms.
% and it's not clear that making this distinction is beneficial 
%
Indeed, PLs are designed to support one or many paradigms, and 
recent PLs are often explicitly designed to 
take the best from an ingenious mix of paradigms.
%Therefore,
%for PLs not only a family contains many species, but also a species
%may belong to many families, differently from what happens in
%biology. 
On the other hand, 
while there is no individual variability at the level of species, the
situation is very different at the level of paradigms: new paradigms
emerge (speciation), they compete (selection) and they often merge
(hybridise). Moreover, for some multi-paradigm language the different
paradigms are somehow orthogonal, such as in Scala;
% e.g., the functional and the OO styles in the Scala language. T
therefore 
the mix of paradigms is not just a matter of hybridisation, but an
actual overlapping of classes.%, with no biological counterpart.

Interestingly, a biological counterpart to this scenario is
represented by the so called multilevel genealogical discordance, that
is, when
the pattern of phylogeny at one level of the biological hierarchy
fails to map onto patterns at other levels. Not surprisingly, such a
discordance is especially notable for microbial organisms, where the
extensive presence of lateral gene transfer, hybridisation, and
recombination lead to reticulate phylogenies. 
Multilevel lineages have been advocated by Matt Haber \cite{Haber} 
to let phylogeny reconstructions take into account that biological
gradients also apply over levels of the hierarchy.
%, in addition to the more familiar gradients that extend over time
%and space. 
Moreover, as observed by 
Haber, multilevel discordance is related to the lineages
\emph{multiple decomposability} problem, that is reminiscent of the
problem of partitioning PLs into programming paradigms mentioned
above.

We leave the case of paradigms as a further open problem.
We argue that object-orientation is just an instance of this problem,
that can be better addressed in a 
\emph{multilevel evolution framework}
rather than resorting to convergent evolution.
In this view, we conclude pointing out an intriguing conjecture, that
might shed light on the multilevel explanation. At the level of
species the evolution of PLs is powerfully driven by syntactic
traits, whereas at the level of programming paradigms
the evolution might
be driven by ``semantics/behavioural traits'' corresponding to
different ways to encode a behaviour. In the functional paradigm
a behaviour is encoded as a function, in the imperative paradigm it is
represented as a sequence of steps, in the declarative paradigm a
behaviour is a property to be satisfied and in the OO paradigm 
it is encoded as an abstract data type. Syntactic and semantic traits
are clearly related but the study of their interplay might give
insights about the hierarchical relationship between PLs and
programming paradigms.

\section{Rich evolutionary patterns as explanatory tools}

%Darwinian theory is much more than random variation and natural
%selection....discuss Lane's paper.
%Discuss rich patterns.
%\medskip\noindent

%{\bf ...qui ci sta bene qualcosa sul fatto che la Teoria \`e un
%  'Programma di Ricerca' e che nella fascia esterna contiene numerosi
%  pattern...} 
The Darwinian Theory of Evolution is much more than random variation
and natural selection, it offers a number of sophisticated
evolutionary patterns that provide for interesting and useful
explanatory tools. While in the previous section we discussed the
basic building blocks of the evolutionary framework, in this section
we show that richer patterns in the evolutionary research program
can be recognised also in the context of programming languages. 

\paragraph{Co-evolution.} By analogy with the co-evolution of human
language and brain, PLs have clearly co-evolved with hardware
technology. 
Besides the two radiation events described in \cite{SoleValverdePLEvo}
corresponding to the birth of structured programming in the 1950s-60s
and the personal computer revolution in the 1980s, we can identify a
number of recent technological innovations that determined major
evolutionary leaps in mainstream programming languages
\cite{CrafaJLAMP}. 
The first one is the advent of the Internet, and especially its appeal
to the market, which shifted the PL goals from efficiency to
portability and security an  also promoted scripting PLs,
such as JavaScript and PHP, to write 
programs to be embedded into web pages and web servers. Moreover, the
fact that nowadays efficient hardware can only be parallel (by means
of multicores and GPU processors but also clusters of machines),
boosted PLs that support parallel and distributed programming, up to
Cloud computing. Finally, the smart technologies provided by the
Internet of Things increase the huge amount of data that can be
collected, and ask for PLs that support the so-called High Performance
Computing needed to deal with the Big Data era.

Interestingly, in the case of programming languages we can devise
another important co-evolving lineage, that has no biological
counterpart: the
advances of theoretical research. Indeed, mainstream programming
and theoretical research on PLs have been mutually influenced:
Robin Milner's Turing lecture \cite{Milner} recalls that suitable
programming abstractions come from a dialectic between the
experimental tests conducted by practical programming and the deep
mathematical tests conducted by the theoretical approach. 
The formal languages studied by theoreticians are well suited to test
new programming primitives and new mix of ``language traits'' in a
concise and expressive model. In other terms, they allow for
\emph{experimentation in a controlled environment}, thus testing and
promoting language mutations that are not necessarily driven by the
actual environment or the short-term future.
Even if also in biology mutations are not (always) driven by
adaptation, such a designed testing and experimentation
has no equal both in biology and human language evolution. 
Experimental manipulations can be conducted in some cultural systems,
especially in the technological systems, but the possibility of a
direct and strict interaction between the theoretical research and the
programming practice is distinctive of PLs.

\paragraph{Macro-evolutionary trends.}
A macro-evolutionary trend is a transversal development that
encompasses different species. In the realm of PLs we can recognise a
macro-trend increasing the abstraction level provided by
languages. Indeed, new languages provide support for more declarative
programming, focusing on ``what to do'' rather than on ``how to do
it'' \cite{CrafaJLAMP}. The details of the implementation of ``how to
do it'' are progressively moved under the hood by increasing the
complexity of the language runtime; for instance consider Java's
automatic garbage collector as opposed to C++'s fine-grained, powerful
but error-prone, control over the memory.

This evolutionary trend is explained by the fact that higher level
programming abstractions enhance program correctness and productivity,
but it is important to observe that it is achievable because of the
underlying (co-evolving) trend that provides for increasingly
efficient hardware which supports stratification of virtual machines
and increasingly complex runtime systems.

\paragraph{Niche construction.} We have already observed that modern
PLs are designed as a mix of programming paradigms. Moreover, modern
software systems, such as those distinctive of innovative
Internet-driven companies such as Google, Facebook, LinkedIn, are
written using a mix of languages, actually a mix of software layers
provided by different language frameworks, that interoperate at
different abstraction levels. 

On the other hand, we can identify a specific \emph{ecosystem of
  languages} for the Web development. Rich and dynamic websites, like
Twitter's, Amazon's or eBay's websites, involve the development of a
back-end, connected to a database, and a front-end for user's
interactions. In particular, the languages 
involved in the front-end development, that is, HTML5 to deal with the
page content, CSS to deal with the page appearance and JavaScript to
deal with the page behaviour, establish a real niche-construction
effect: they are different languages but they mutually affect their
evolution. 

\paragraph{Exaptation.}
An interesting example of functional shift can be identified observing
that after fifty years of functional programming languages, the
distinctive traits of those languages, that is functions, shine in new
languages essentially because they leverage effective concurrent
programming. Indeed, for a long time functional programming techniques
had been confined to languages that have never become mainstream. But
looking at a function as an abstraction that represents a behaviour,
which can be passed around and composed, allows for a smooth
integration with the design of the concurrent execution of different
tasks. Moreover, the spatial thinking supported by functional
programming smoothly fits object-oriented programming's ability of
structuring software systems. Therefore well-established mainstream
imperative (an object-oriented) languages such as C++ and Java
recently embarked on a deep change to introduce higher-order functions
(in C++11 and Java8) that leverage efficient parallel programming over
large data structures. 

\section{Conclusions}

The programming languages development represents a well structured 
case study to investigate how deep is the analogy between 
biological and cultural evolution. It is supported by a rich
and complete fossil record and lays in the intersection between two
notable streams of work in the realm of cultural evolution: the
evolution of human languages and the evolution of technology. 

In this paper we argued that in order to understand how much the
Evolutionary Theory can provide for an explanatory
framework in this realm, it is important to carry out 
a precise assessment of how far the basic ingredients of the
Darwinian evolution can be rephrased to deal with PLs.
We showed that many evolutionary mechanisms, such as diversification
processes and external pressures have an actual
correspondent. However, we raised a number of critical issues, such as
the identification of replication mechanisms and the lack of
populational thinking entailed by the species definition. 
The analysis of PLs phylogenies put forward hierarchical
considerations, 
and we suggested the use of a multilevel evolution framework to
encompass the case of programming paradigms. Finally, we showed that
richer evolutionary patterns, such as co-evolution, macro-evolutionary
trends, niche construction and exaptation, can be effectively applied
to describe and interpret the complex evolution of programming
languages. 

This paper fits in the recent debate emphasising the disanalogiges
between the way variation, selection and inheritance operate in the
biological and cultural cases. We have seen PLs as a bridge between
the evolution of technology and that of human languages; 
an interesting future step will be comparing the PLs case with
the study of patented inventions, which share with PLs the highly
multi-parental genealogy and the complex and reticulated, possibly
multilevel, lineages.  
Finally, a distinctive feature also worthy of further
investigation is the impact of theoretical research on the
evolution of PLs as a source of %(even possibly retrospective)
experimental manipulation. 

%We discussed
% what extent the evolutionary mechanisms distinctive of biology can be applied to this area,

%since it encompasses both the tangled/complex/involved and
%fluid/variable nature of cultural systems that embody the
%difficulties due to social effects, but also the the preciseness of a
%(almost) rigorous scientific/formal account distinctive of computer
%science. Notice that everything is known about the structure of a PL
%and of the computer hosting a running program, but this does not mean
%that the behaviour of a program is fully deterministic not to say
%correct w.r.t. its specification/design. The (formal) analysis and
%verification of software systems is indeed a major branch of (research
%in) computer science. So the problem under consideration might be more
%disciplined than other cultural systems, but still very complex.

%What is the interest of studying evolutionary patterns in PLs?

%Evolutionary account can reveal/explicitate interactions between
%different development and influence mechanisms.

%Evolutionary patterns can be used as explanation tools, not to force
%the evolutionary account but tho enrich a 
%frame of a ``pluralism of explatantions''

%....population thinking,  multilevel evolution, experimental
%manipulations. ...

%\section*{References}

\end{document}

\end{document}
%%%%%%%%%%%%%%%%%%%%%%%%%%%%%%%%%%%%%%%%%%%%